\documentclass[twocolumn,amsmath,amssymb,showpacs,english,aps]{revtex4}

\usepackage{graphicx}
 
\makeatletter
\usepackage{times}
\usepackage[T1]{fontenc}
\usepackage[latin1]{inputenc}
\usepackage{amssymb}
 
\makeatletter
\numberwithin{equation}{section} 
\numberwithin{figure}{section} 
\newtheorem{theorem}{Theorem}
\newcommand{\be}{\begin{equation}}
\newcommand{\ee}{\end{equation}}
\newcommand{\ben}{\begin{eqnarray}}
\newcommand{\een}{\end{eqnarray}}

\newcommand{\nd}{\noindent}

\usepackage{babel}
\makeatother
\begin{document}
 
\title{Power-law  random walks}

\author{$^1$C. Vignat and A. Plastino$^2$}
\affiliation{$^1$ L.P.M., E.P.F.L, Lausanne, Switzerland \\ $^2$
La Plata National University, Exact Sciences Faculty\\ $\&$
National Research Council (CONICET) \\ C. C. 727 - 1900 La Plata -
Argentina  }
 
\pacs{05.40.Fb, 05.40.Jc, 05.40.-a, 02.50.-r}

\begin{abstract}
\nd We present some new results about the distribution of a random
walk whose independent steps follow a $q-$Gaussian distribution with
exponent $\frac{1}{1-q};\,\,\,q \in \mathbb{R}$. In the case
$q>1$ we show that a stochastic representation of the point
reached after $n$ steps of the walk can be expressed explicitly
for all $n$. In the case $q<1,$ we show that the random walk can
be interpreted as a projection of an isotropic random walk, i.e. a
random walk with fixed length steps and uniformly distributed
directions.
\end{abstract}
\maketitle
 
\section{Introduction}
\nd The name ``random walk" (RW) was originally proposed by
Pearson in 1901 \cite{pearson} with reference to a simple model to
describe mosquito infestation in a forest,
although previous important work in related subjects had been
already published by  Lord Rayleigh.
 Around the same time, the theory of
 random walks was also developed by Bachelier in a remarkable doctoral thesis
 published in 1900 \cite{pearson}. He  proposed the random walk as the fundamental
 model for financial time series, long before this idea became the
 basis for modern theoretical finance. He also made the connection between
 discrete random walks and the continuous diffusion (or heat) equation,
 which is a major scientific theme. Around the same time as Pearson' s work, Einstein also
 published his seminal paper on Brownian motion (normal diffusion), a random walk driven by
  collisions with gas molecules. Similar theoretical ideas were
   also published independently by Smoluchowski \cite{pearson}.
   One can confidently assert that RW is one of the most basic themes
   of science.
 
\nd
 The statistical properties of random walks tend toward universal distributions
  after large numbers of independent steps. In the case of the
  concomitant probability distribution function (PDF) for
   the final position, the result for isotropic random walks is a
   multidimensional generalization of the Central Limit Theorem (CLT)
  for  sums of independent, identically distributed (IID) random variables. When
the assumptions of the  Central Limit Theorem break down, random
walks can exhibit rather  different behavior from that of normal
diffusion.   For instance, the limiting distribution for the
position of a Brownian   particle  may not be Gaussian. In
particular, power-law distributions become of
paramount importance in such a context. One
 way to violate the CLT with IID displacements is via ``heavy-tailed"
 probability distributions, which assign sufficient probability
 to very large steps so that the variance is infinite. In this
 context one speaks of anomalous diffusion (AD). In an AD-scenario
 power-law probability distributions and power-law
 entropies become ubiquitous. The associated literature is really
 vast. See for instance, Ref. \cite{okamoto,gellmann,lissia,euro,anteneodo}
 and references therein.
 
 \nd In this effort we revisit the distribution of a random
walk whose independent steps follow a power-law distribution of the $q-$Gaussian
type with
exponent equal to $1/(1-q);\,\,\,q \in \mathbb{R}.$ 
In the case $q>1$ we show
that a stochastic representation of the point reached after $n$
steps of the walk can be expressed explicitly for all $n$ so that the superstatistics
framework holds even in the anomalous diffusion case. In the
case $q<1,$ we show that the $q-$Gaussian random walk can be interpreted as a
projection of an isotropic random walk, i.e. a random walk with
fixed length steps and uniformly distributed directions.

 \section{The case $q>1$}
 
\nd A random vector $X\in\mathbb{R}^{p}$ is of the  q-Gaussian kind  if its probability density is
\begin{equation}
f_{X}\left(X\right)=Z_{q}^{-1}
\left(1+X^{t}\Lambda^{-1}X\right)^{\frac{1}{1-q}}
\label{eq:q>1gaussian}
\end{equation}
where the number of degrees of freedom $m$, dimension $p$ and
non-extensivity parameter $q$ are related as
\begin{equation}
m=\frac{2}{q-1}-p,\label{eq:mq>1}
\end{equation}
This distribution has finite  covariance matrix $K=EXX^{t}$ provided that $m>2$ or equivalently $q<\frac{p+4}{p+2}.$ In that case, the covariance matrix is related to  the scaling matrix  $\Lambda$ in the fashion
\begin{equation}
\Lambda=\left(m-2\right)K.\label{eq:Kq>1}
\end{equation}
Moreover,
the partition function $Z_q$ reads
\[
Z_{q}
= \left ( \frac{\Gamma\left(\frac{1}{q-1}\right)}
{\Gamma\left(\frac{1}{q-1}-\frac{p}{2}\right)\vert\pi\Lambda\vert^{1/2}}  \right ) ^{-1}.
\]
Note that the usual Gaussian distribution corresponds to the limit case $q\rightarrow 1^{+}.$
\nd We recall \cite{3NC,vignat,plastino} that random vector $X$
can be expressed as the Gaussian scale mixture 
\footnote{a random vector U is a Gaussian scale mixture if $U=\sqrt{b}G$ where $G$ is a Gaussian vector and $b$ is a
random variable {\it independent} of $G$}
\begin{equation}
X=\frac{\Lambda^{1/2}G}{\sqrt{a}}\label{eq:representation_q>1}\end{equation}
where $G$ is a $p-$variate, unit covariance Gaussian random vector and $a$ is a
random variable independent of $G$ that follows a $\chi^{2}$
distribution with $m$ degrees of freedom. Representation
(\ref{eq:representation_q>1}) reflects exactly the notion of
superstatistics as introduced by Beck and Cohen
 \cite{B1,B2,B3,B4,B5,B6,B7}: a $q-$Gaussian random system with
   $q>1$ can be interpreted as a Gaussian system submitted to
multiplicative fluctuations following an inverse chi distribution.
 
\subsubsection{the finite covariance case}
\nd In the context of a random walk, we are interested in the
distribution of the normalized random vector
\begin{equation}
Z_{n}=\frac{1}{\sqrt{n}}\sum_{i=1}^{n}X_{i}
\label{eq:randomwalk}
\end{equation}
where random vectors $X_{i} \in \mathbb{R}^{p}$ are independent and q-Gaussian
distributed according to (\ref{eq:q>1gaussian}), each with $m$
degrees of freedom and covariance matrix $K.$ Random vector
$Z_{n}$ can be characterized by the following
 
 \begin{theorem}
 \label{thm:stochastic}
A stochastic representation of random vector $Z_{n}$ is
\begin{equation}
\label{Zn} 
Z_{n} \sim
\alpha_{m,n}\left(\sum_{i=1}^{n}\frac{1}{\nu_{i}}\right)^{\frac{1}{2}}X
\end{equation}
where the $p-$variate vector $X$ is q-Gaussian distributed with $nm$
degrees of freedom and covariance matrix $K,$ where the constant
quantity  $\alpha_{m,n}=\frac{1}{n}\sqrt{\frac{m-2}{m-\frac{2}{n}}}$, while
$\left\{ \nu_{1},\dots, \nu_{n-1}\right\} $ are Dirichlet
distributed \footnote{Vector $\left(v_{1},\dots,v_{n-1}\right)$ has
a Dirichlet distribution with parameters
$\left(\alpha_{1},\dots,\alpha_{n}\right)$ if its distribution has
density \[
f\left(v_{1},\dots,v_{n-1}\right)=c.v_{1}^{\alpha_{1}-1}\dots
v_{n-1}^{\alpha_{n-1}-1}\left(1-v_{1}-\dots-v_{n-1}\right)^{\alpha_{n}-1}\]
over the $n-$dimensional simplex $\sum_{i=1}^{n-1}v_{i}<1, \,\, v_{i}\ge0.$ } with
parameters $m_{1}=\dots=m_{n}=m$ and independent of vector $X$.
\end{theorem}
 
\nd {\bf Proof:} We follow here the proof given in \cite{Dickey}:
a linear combination of Gaussian scale mixtures is itself a
Gaussian scale mixture since \cite{symbol} \ben &
Z_{n}=\frac{1}{\sqrt{n}}\sum_{i=1}^{n} \frac{\Lambda^{1/2}G_{i}}
{\sqrt{a_{i}}}\cr & \sim \frac{1}{\sqrt{n}}
\sqrt{\sum_{i=1}^{n}\frac{1}{a_{i}}} \Lambda^{1/2}G  \cr &
=\frac{1}{\sqrt{n}}
\sqrt{\sum_{i=1}^{n}\frac{\sum_{j=1}^{n}a_{j}}{a_{i}}}
\frac{\Lambda^{1/2}G}{\sqrt{\sum_{j=1}^{n}a_{j}}}\een where $G$ is
a $p-$variate Gaussian vector with unit covariance. Now we remark
that random variables $a_{i}$ are chi-square distributed with $m$
degrees of freedom; thus, by Luckacs' result, each random variable
\begin{equation}
\nu_{i}=\frac{a_{i}}{\sum_{j=1}^{n}a_{j}}\label{eq:dirichlet}\end{equation}
is independent of $\sum_{j=1}^{n}a_{j}.$ Moreover, equality
(\ref{eq:dirichlet}) shows that random variables $\left\{
\nu_{i}\right\} _{1\le i\le n-1}$ are Dirichlet distributed. At
last, since $\sum_{j=1}^{n}a_{j}$ is chi-square distributed with
$mn$ degrees of freedom, we deduce that
\[
\frac{\sqrt{nm-2}}{\sqrt{m-2}}\frac{\Lambda^{1/2}G}{\sqrt{\sum_{j=1}^{n}a_{j}}}\]
is a $q-$Gaussian vector with covariance matrix $K$ and
$mn$ degrees of freedom.\,\,$\square$
\\
 
\nd A striking result is thus obtained: at its $n-$th step, a
q-Gaussian random walk with $q>1$ is a scale mixture of a q-Gaussian
vector. We note that this property {\it holds true for any random
walk with independent steps following a Gaussian scale mixture}. The
fact that this property extends to q-Gaussian distributions is
indeed remarkable. As described in the preceding proof, this special behavior is a
consequence of a famous result by Lukacs \cite{lukacs} about the
Gamma distributions, which are precisely the ones that rule the
fluctuations described by the superstatistics theory
\cite{B1,B2,B3,B4,B5,B6,B7}. 
\\
Moreover, the
non-extensivity parameter $q'$ of vector $X$ in (\ref{Zn}) is related to the
parameter $q$ of each step as \be q' =
1+\frac{2(q-1)}{2+m(n-1)(q-1)}. \ee We note that the dimension $p$ of the random walk does not appear in this formula.
 
\nd The curves in Figure 1 represent $q'$ as a function of $n$ for
\begin{enumerate} \item  $m=5$ and $q=5, 3$ and $1.5$ (top to bottom) on the three top
curves; \item for $m=10$ and $q=5, 3$ and $1.5$ (top to bottom) on
the three middle curves; \item for $m=30$ and $q=5, 3$ and $1.5$
(top to bottom) on the three lower curves. \end{enumerate}

\begin{figure}[htbp] 
   \centering
   \includegraphics[scale=0.45]{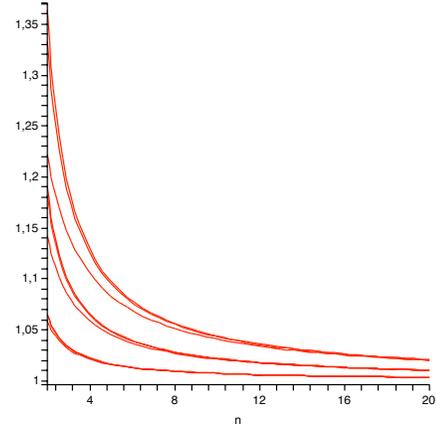}
   \caption{$q'$ as a function of $n$ for $m=5$
   and $q=5, 3$ and $1.5$ (top to bottom)
   on the three top curves; for $m=10$ and $q=5, 3$ and $1.5$
   (top to bottom) on the three middle curves;
   for $m=30$ and $q=5, 3$ and $1.5$ (top to bottom) on the three lower curves}
   \label{fig:example}
\end{figure}
 
\nd These curves confirm the three following results:
\begin{enumerate}
 
\item since the variance is finite, the central limit theorem
applies and $Z_{n}$ converges to a Gaussian vector
with covariance matrix $K$, and thus $q'$ converges to $1$
\item the convergence to a Gaussian vector
is all the faster since the number of degrees of freedom $m$
is large - or equivalently, since the independent steps $X_{i}$ are closer to Gaussian steps
\item for a large enough value of $m$, the convergence process of $q'$ to $1$
is relatively insensitive to the value of $q.$
\end{enumerate}
 
\nd Unfortunately, the probability density for the scaling random
variable  \be  U_{m,n} =
\alpha_{m,n}\left(\sum_{i=1}^{n}\frac{1}{\nu_{i}}\right)^{\frac{1}{2}},
\ee
can not be  explicitly given.  Figures 2 and 3 below depict  an
estimation of the probability distribution function  for  $U_{m,n}$
after $n=5, 10, 15, 20$ and $30$ steps of the
random walk in the cases $m=5$ and $m=25$.
Note that different
scales have been employed. These figures  clearly  exhibit the
convergence of the random variable $U_{m,n}$ to the deterministic
unit-constant, as required by the central limit theorem.

\begin{figure}[htbp] 
   \centering
   \includegraphics[scale=0.25]{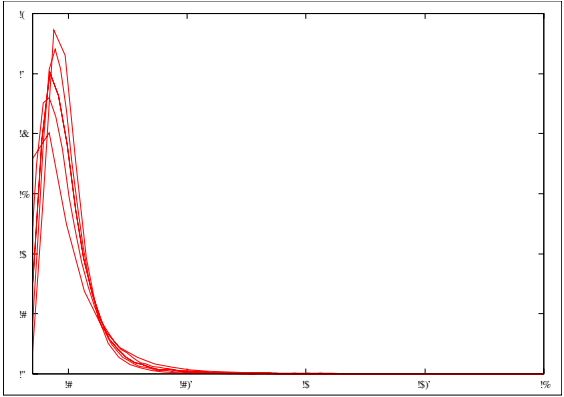}
   \caption{estimated pdf of $U_{5,n}$ with $n=5, 10, 15, 20$ and $30$
   (bottom to top)}
   \label{fig:example}
\end{figure}
 
\begin{figure}[htbp] 
   \centering
   \includegraphics[scale=0.25]{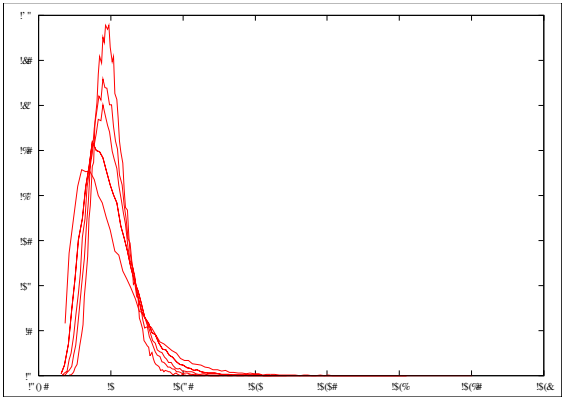}
   \caption{estimated pdf of $U_{25,n}$ with $n=5, 10, 15, 20$ and $30$
   (bottom to top)}
   \label{fig:example}
\end{figure}
 
\subsubsection{The infinite covariance case:  L\'{e}vy flights}
\nd Let us assume now $m<2$ so that each of the steps of the random
walk has infinite covariance. Let us consider the unnormalized random
walk
 
\begin{equation}
Z_{n} = \sum_{i=1}^{n} X_{i}
\label{eq:unnormalized}
\end{equation}
 
The distribution of any component $X_{i}^{(k)}$ of vector $X_{i}$,
\[
f_{x}(x) = \frac{\Gamma(\frac{m+1}{2})}{\Gamma(\frac{m}{2})\sqrt{\pi \Lambda_{k,k}}} (1+\frac{x^{2}}{\Lambda_{k,k}})^{-\frac{m+1}{2}}
\]
behaves as \be f_{x}(x)
\sim \vert x\vert ^{-m-1}, \,\, \vert x \vert \sim \infty \ee so that  the
number of degrees of freedom $m$, for  $m\le2$, coincides with the
L\'{e}vy index of the $q-$Gaussian distribution.
Now,  by direct
application of the L\'{e}vy-Gnedenko central limit theorem
(\cite{gnedenko,milan}) one immediately realizes  that \be
\label{thm:LevyGnedenko} \frac{1}{n^{1/m}}\Lambda^{-1/2} Z_{n} \Rightarrow S_{m},
\ee where $S_{m}$ denotes a vector, each components of which follows a symmetric alpha-stable distribution
with L\'{e}vy index $m$.
 
\nd A quite  interesting result worth quoting at this point is that,
although the involved variables have infinite covariance, the
superstatistics principle still applies in the following fashion:
 
\begin{theorem}
For all $n,$ the normalized random walk $\frac{1}{n^{1/m}}Z_{n}$ is
distributed as a  Gaussian scale mixture.
Further,  the distribution of the mixing variable converges, as
$n\rightarrow +\infty$, to a stable distribution with L\'{e}vy index
$m/2$.
\end{theorem}
 
\nd {\bf Proof:} We use the first part of the proof of Th.
\ref{thm:stochastic}: \be Z_{n}=\sum_{i=1}^{n} \frac{\Lambda^{1/2} G_{i}}
{\sqrt{a_{i}}} \sim \sqrt{\sum_{i=1}^{n}\frac{1}{a_{i}}} \Lambda^{1/2} G \ee One
can easily check that each random variable $\frac{1}{a_{i}}$ has
L\'{e}vy index $m/2$. Now, the L\'{e}vy-Gnedenko theorem yields
\[
\frac{1}{n^{2/m}} \sum_{i=1}^{n}\frac{1}{a_{i}} \Rightarrow S_{m/2},
\]
so that
\[
\frac{1}{n^{1/m}} Z_{n} \Rightarrow \sqrt{S_{m/2}} \Lambda^{1/2} G.
\]
$\square$
\\
Note that this result is coherent with the representation
(\ref{thm:LevyGnedenko}) as given by the L\'{e}vy Gnedenko theorem.
Indeed, according to a classical result about stable random
variables (see \cite{feller}), if $S_{\alpha}$ and $S_{\alpha'}$
denote two independent stable random variables with respective
indices $\alpha$ and $\alpha'$ then,
\[
S_{\alpha} S_{\alpha'}^{1/\alpha} \sim S_{\alpha''}
\]
with $\alpha''=\alpha \alpha'$. In the  above situation, this result applies component-wise with $\alpha=2$
and $\alpha'=m/2.$

\section{Case $q<1$}
 
\nd The $p-$variate $q-$Gaussian distribution in the case $q<1$ writes
explicitly as\begin{equation}
f_{X}\left(X\right)=Z_{q}^{-1}\left(1-X^{t}\Lambda^{-1}X\right)_{+}^{\frac{1}{1-
q}} \label{eq:q<1gaussian}\end{equation}
with notation $\left(x\right)_{+}=\max\left(x,0\right).$ The
covariance matrix of vector $X$ is finite and writes $K=EXX^{t}=d^{-1}\Lambda$ with
$d=p+2\frac{2-q}{1-q}$.
Moreover, the partition function $Z_{q}$
writes
\be
Z_{q} = \left ( \frac{\Gamma\left(\frac{2-q}{1-q} +
\frac{p}{2}\right)}
{\Gamma\left(\frac{2-q}{1-q}\right)\vert\pi\Lambda\vert^{1/2}}
\right ) ^{-1}.
\ee
A stochastic representation of a vector $X$
following this distribution is
\be
\label{eq:stochastic}
X=\frac{\Lambda^{1/2}G}{\sqrt{G^{t}G+b}},
\ee where the random
variable $b$ is hi-square distributed with
$2\frac{2-q}{1-q}$ degrees of freedom and independent of the $p-$variate, unit variance
Gaussian vector $G$ and c. 
\nd Let us consider now the
random walk 
\begin{equation}
Z_{n}=\frac{1}{\sqrt{n}} \sum_{i=1}^{n}X_{i}, 
\label{Znq<1}
\end{equation}
where the random vectors $X_{i}$
are independent and follow distribution
(\ref{eq:q<1gaussian}).
 
\vskip 3mm
 
\nd Although, contrarily to the case $q>1$, no explicit stochastic
representation can be provided for (\ref{Znq<1}),
 this kind of random walk can be given an interesting interpretation,  as follows:
 
\begin{theorem}
If $Y_{n}$ is an isotropic
$d-$dimensional random walk \cite{pearson}
\[
Y_{n}=\frac{1}{\sqrt{n}}\sum_{i=1}^{n}T_{i}\] where
$T_{i}\in\mathbb{R}^{d}$ are independent random vectors with unit
length $\Vert T_{i}\Vert=1$ and uniformly distributed direction
\footnote{in other words, each $T_{i}$ is uniformly distributed on the unit
sphere $\mathcal{S}_{p}=\left\{ X\in\mathbb{R}^{p}  \vert  \Vert X\Vert=1\right\} $
} and if
\begin{equation}
2\frac{2-q}{1-q}\in\mathbb{N}
\label{eq:condition}
\end{equation}
then $\Lambda^{-1/2} Z_{n}$ is the $p-$dimensional marginal of $Y_{n}$ with
\begin{equation}
p=d-2\frac{2-q}{1-q}.
\label{eq:p_d}\end{equation}
\end{theorem}
 
\nd {\bf Proof:} A vector $T_{i}$ uniformly distributed on the sphere
$\mathcal{S}_{d}$ has stochastic representation
\[
T_{i} = \frac{\tilde{G}}{\Vert \tilde{G} \Vert}
\]
where $\tilde{G}\in \mathbb{R}^{d}$ is a Gaussian random vector with unit covariance. Thus
stochastic representation (\ref{eq:stochastic}) shows that $\Lambda^{-1/2}X$ is the $p-$variate marginal of $T_{i}$ (see \cite{plastino}) \,\, $\square$
\\
\nd In a physical context, this result can be interpreted as
follows: assume that we observe a $p-$dimensional $q-$random walk
whose nonextensivity parameter $q<1$ verifies condition
(\ref{eq:condition}); then a reasonable hypothesis is that one
observes only a part (some components) of a more natural
higher-dimensional random walk, namely a $d-$variate isotropic
random walk with $d$ defined as in (\ref{eq:p_d}).
\newline
\newline
\nd As an illustration, Figure 4 shows the first five steps of a
(dimension $d=3$)-isotropic random walk with unit length steps,
starting from the origin. The projected random walk in the plane
$z=0$ corresponds to $p=2$ and $q=3$ [according to (\ref{eq:p_d})].
The projected random walk on the $x-$axis is characterized by $p=1$
and thus by $q=+\infty$, which corresponds to the uniform
distribution on the interval $[-1,+1]$.
 
\begin{figure}[htbp] 
   \centering
   \includegraphics[scale=0.6]{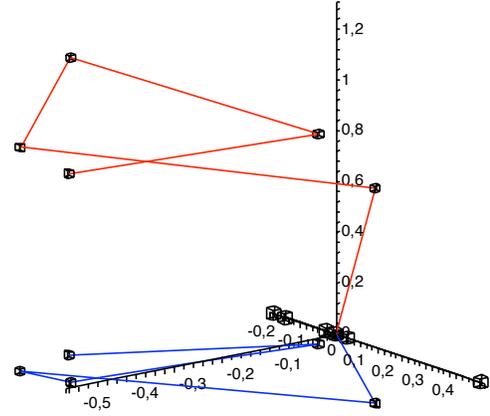}
   \caption{an isotropic random walk in 3 dimensions and its projections}
   \label{fig:example}
\end{figure}
 
\nd Another useful result about the q-Gaussian random walk $Z_{n}$,
as defined by (\ref{Znq<1}) with $q<1$, is given by

\begin{theorem}
If {\rm (i)} $Z_{n} \in \mathbb{R}^{p}$ is a $q-$Gaussian random
walk with $q<1$ and {\rm (ii)}   $\{a_{i}\}_{1 \le i \le n} $ are independent random variables
(independent of all $X_{i}$'s for $1\le i \le n$ in Eq.
(\ref{Znq<1})) that follow a chi-distribution with $d$
degrees of freedom such that
\be d=p+2\frac{2-q}{1-q},
\ee
then the random walk \be \tilde{Z}_{n} = \frac{1}{\sqrt{n}}
\sum_{i=1}^{n} a_{i} X_{i}, \ee is a {\rm Gaussian} random walk with
independent steps, each with covariance $\Lambda$.
\end{theorem}
 
\nd {\bf Proof:} The fact that each step $a_{i} X_{i}$ is a Gaussian
vector is proved in \cite{3NC}. The covariance matrix of $a_{i} X_{i}$
is easily computed as \be E (a X_{i})(a X^{t}_{i}) = Ea^{2}
EX_{i}X^{t}_{i} = d \frac{1}{d} \Lambda = \Lambda \ee and the
independence of the steps results from the assumptions. $\square$

\section{Conclusions}
\nd We have proved here some new results
for random walks governed by distributions of the power-law type. Summing up:

\begin{itemize}
 
\item In the case $q>1$ (Section II) we saw that a stochastic representation of the point
reached after $n$ steps of the walk can be expressed explicitly for
all $n$. 
\item Moreover, Theorem {\bf 2} allows one to highlight the fact that, even in the
 L\'{e}vy (infinite covariance) case, the superstatistics framework remains still
 valid, a rather remarkable result
 
\item In the case $q<1$ (Section III),  we ascertained  that the random walk can be
interpreted as a projection of an isotropic random walk, i.e. a
random walk with fixed length steps and uniformly distributed
directions. 

\item Moreover, Theorem {\bf 4} shows that a $q-$Gaussian
random walk with $q<1$, each step of which being subjected to
independent, multiplicative chi-distributed fluctuations, is exactly
a Gaussian random walk, a fact that can be qualified as a {\it dual
superstatistics}.
\end{itemize}

\acknowledgments
C.V. would like to thank C. Tsallis for useful discussions about stable random walks, and the people at I.R.M.A.R. Library, University of Rennes, for invaluable support.

\end{document}